\documentclass[prb,aps,
nobibnotes,
twocolumn,
10pt,
floatfix,
]{revtex4-1}
\usepackage[normalem]{ulem}
\usepackage{amsmath}
\usepackage{amssymb}
\usepackage{graphicx}
\usepackage{bm}
\usepackage{mathrsfs}
\usepackage{mathtools}
\usepackage{booktabs}
\usepackage{multirow}
\usepackage{siunitx}
\usepackage[caption=false]{subfig}
\usepackage{footnote}
\usepackage{fancyhdr}
\usepackage{cancel}
\usepackage{placeins}

\makeatletter
\fancypagestyle{footmark}{
  \fancyfoot[L]{\footmark}
}
\newcommand\markfoot[1]{\gdef\footmark{#1}\thispagestyle{footmark}}
\makeatother

\newcounter{Figcount}
\newcounter{tempFigure}
\newenvironment{figCaption}{%
    
    \setcounter{tempFigure}{\thefigure}
    \setcounter{figure}{\theFigcount}
    }{%
    \setcounter{figure}{\thetempFigure}
    \stepcounter{Figcount}
    }

\newenvironment{tabCaption}{%
    
    }{%
    }

\def\be{\begin{equation}}
\def\ee{\end{equation}}
\def\bea{\begin{eqnarray}}
\def\bea{\end{eqnarray}}
\def\eea{\end{eqnarray}} 	
\def\bs{\begin{split}}
\def\es{\end{split}}
\def\ni{\noindent}

\def\bi{\begin{itemize}}
\def\ei{\end{itemize}}
\def\f{\frac}

\newcommand{\subfigimg}[3][,]{%
  \setbox1=\hbox{\includegraphics[#1]{#3}}
  \leavevmode\rlap{\usebox1}
  \rlap{\hspace*{0pt}\raisebox{\dimexpr\ht1-1\baselineskip}{#2}}
  \phantom{\usebox1}\phantom{1cm}
}

\begin{document}

\title{{\Huge Long-range rhombohedral-stacked graphene through shear}}

\author{ Jean Paul Nery$^{1,2}$, Matteo Calandra$^{3,4}$, and Francesco Mauri$^{1,2,\ast}$ }

\begin{abstract}
\ni \textbf{
The discovery of superconductivity and correlated electronic states in the flat bands of twisted bilayer graphene has raised a lot of excitement. Flat bands also occur in multilayer graphene flakes that present rhombohedral (ABC) stacking order on many consecutive layers. Although Bernal-stacked (AB) graphene is more stable, long-range ABC-ordered flakes involving up to 50 layers have been surprisingly observed in natural samples. Here we present a microscopic atomistic model, based on first-principles density functional theory calculations, that demonstrates how shear stress can produce long-range ABC order.
A stress-angle phase diagram shows under which conditions ABC-stacked graphene can be obtained,
providing an experimental guide for its synthesis.\\
}
\end{abstract}
\maketitle

Multilayer graphene exhibits two main types of stacking.
In Bernal-stacked multilayer graphene (BG) layers are stacked repeatedly in the AB sequence, while in  rhombohedral-stacked multilayer graphene (RG) the stacking is ABC.
The main interest  in RG stems from its flat bands close to the Fermi energy \cite{Pierucci2015,Pamuk2017}, which could lead to exciting phenomena such as superconductivity \cite{Kopnin2013, Munioz2013}, charge-density wave or magnetic orders \cite{Baima2018}. The extent of the flat surface band in the Brillouin zone and the number of electrons hosted increases with  the number of consecutive ABC-stacked layers (saturating at approximately 8 layers) \cite{Koshino2010, Xiao2011, Pamuk2017}. Thus, mastering the thickness of ABC flakes is a way to taylor correlation effects. However, RG is much less common than the energetically favored BG phase \cite{Laves1956} and does not appear isolated\cite{Roviglione2013,Xu2015,Kim2015}. While superconductivity has already been measured in twisted bilayer graphene \cite{Herrero2018,Yankowitz2019,Lu2019}, work on RG has been slower because of the inability to consistently grow or isolate large single crystal samples.

\markfoot{\vspace*{-7mm}\hrule\vspace*{1mm}\sf
\footnotesize{$^{1}$Graphene Labs, Fondazione Istituto Italiano di Tecnologia, Via Morego, I-16163 Genova, Italy. $^{2}$Dipartimento di Fisica, Universit\`a di Roma La Sapienza, Piazzale Aldo Moro 5, I-00185 Roma, Italy.  $^{3}$Sorbonne Universit\'e, CNRS, Institut des Nanosciences de Paris,
UMR7588, F-75252, Paris, France. $^{4}$Department of Physics, University of Trento, Via Sommarive 14, 38123
Povo, Italy. $^\star$e-mail:francesco.mauri@uniroma1.it}}

X-ray diffraction experiments\cite{Lipson1942,Laves1956} have shown that some natural samples contain small amounts of rhombohedral graphite. However, such experiments have not determined if the stacking is random, or if there are many consecutive layers of ABC-stacked graphene, namely, if there is a phase separation between BG and RG.
With these same limitations, also using X-ray diffraction, it was qualitatively noticed 
that shear strain increases the percentage of rhombohedral inclusions in Bernal graphite \cite{Laves1956}.
Ref.~\cite{Boehm1955} proposed a gliding mechanism, but it involved going through an intermediate AA stacking, a high-energy state. Then, ref.~\cite{Laves1956} proposed gliding that avoided AA stacking and involved a shorter displacement.
This pinpointed the gliding mechanism that produces RG, but did not explain the precise nature of the stacking. 

Definitive experimental evidence of long-range ABC order has only been obtained in the last years. After applying shear to BG, over 10 consecutive layers of RG were first observed using selected-area electron diffraction\cite{Lin2012}. More than 14 layers \cite{Balseiro2016, Henck2018}, and up to 50 layers of RG \cite{Mishchenko2019a,Mishchenko2019b} have been observed in exfoliated samples as well. Notice that for a random stacking,
the probability of obtaining $N$ consecutive layers of RG is $1/2^{N-2}$, which corresponds to only 0.02 \% for $N=14$, and becomes extremely small for $N \sim 50$. Thus, there must be some underlying reason, either energetic or kinetic, that explains why this happens.

Here, we propose a mechanism to produce long-range RG staking from BG using shear stress. In particular, we use an atomistic model, based on first-principles calculations, to obtain a stress-angle phase diagram that identifies the conditions for the formation of RG. The required stress 
is similar to that already realized in friction experiments of graphene \cite{Dienwiebel2004,Liu2012b}.

\begin{figure*}[!ht]
\centering
\subfigimg[width=0.8\textwidth]{\textbf{a}}{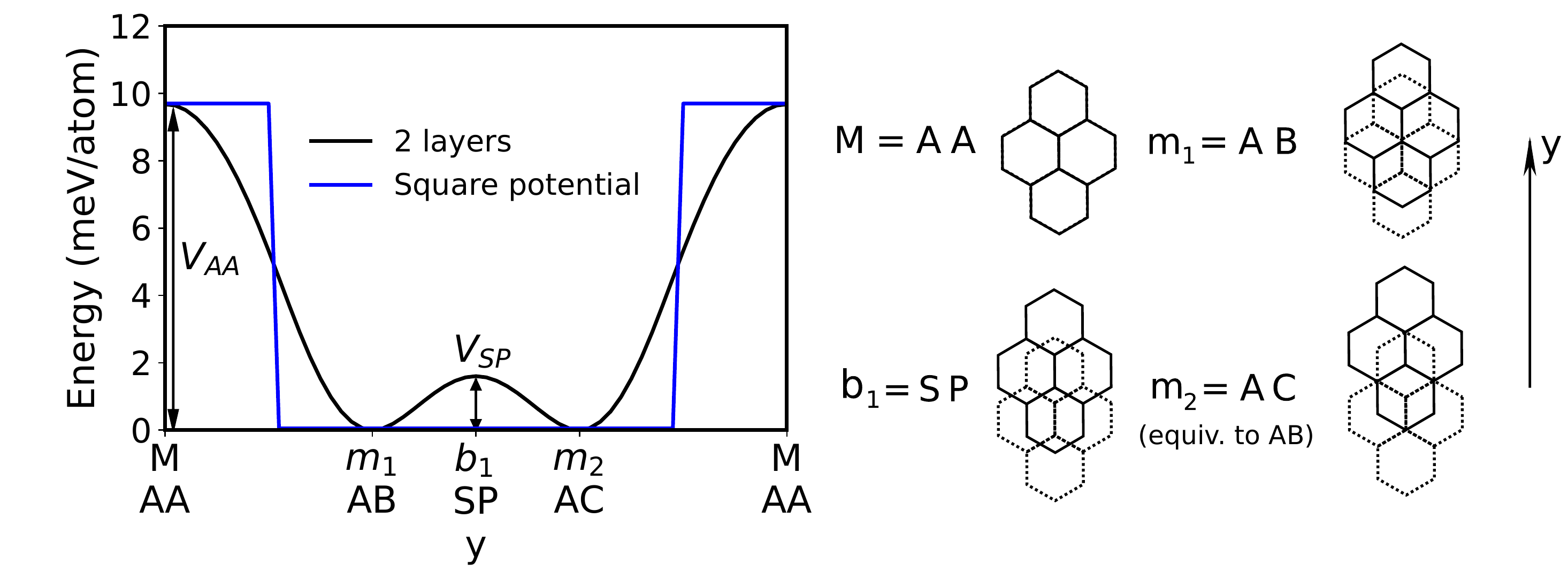} \\
\subfigimg[width=0.8\textwidth]{\textbf{b}}{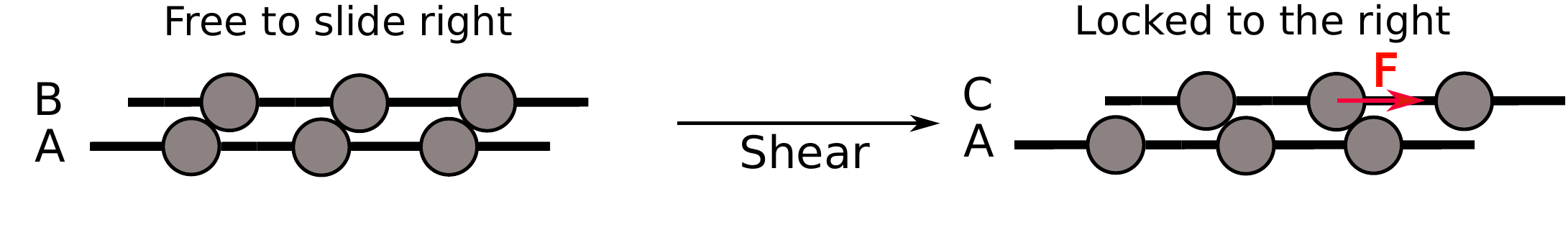}
\vspace{1cm}
\subfigimg[width=0.8\textwidth]{\textbf{c}}{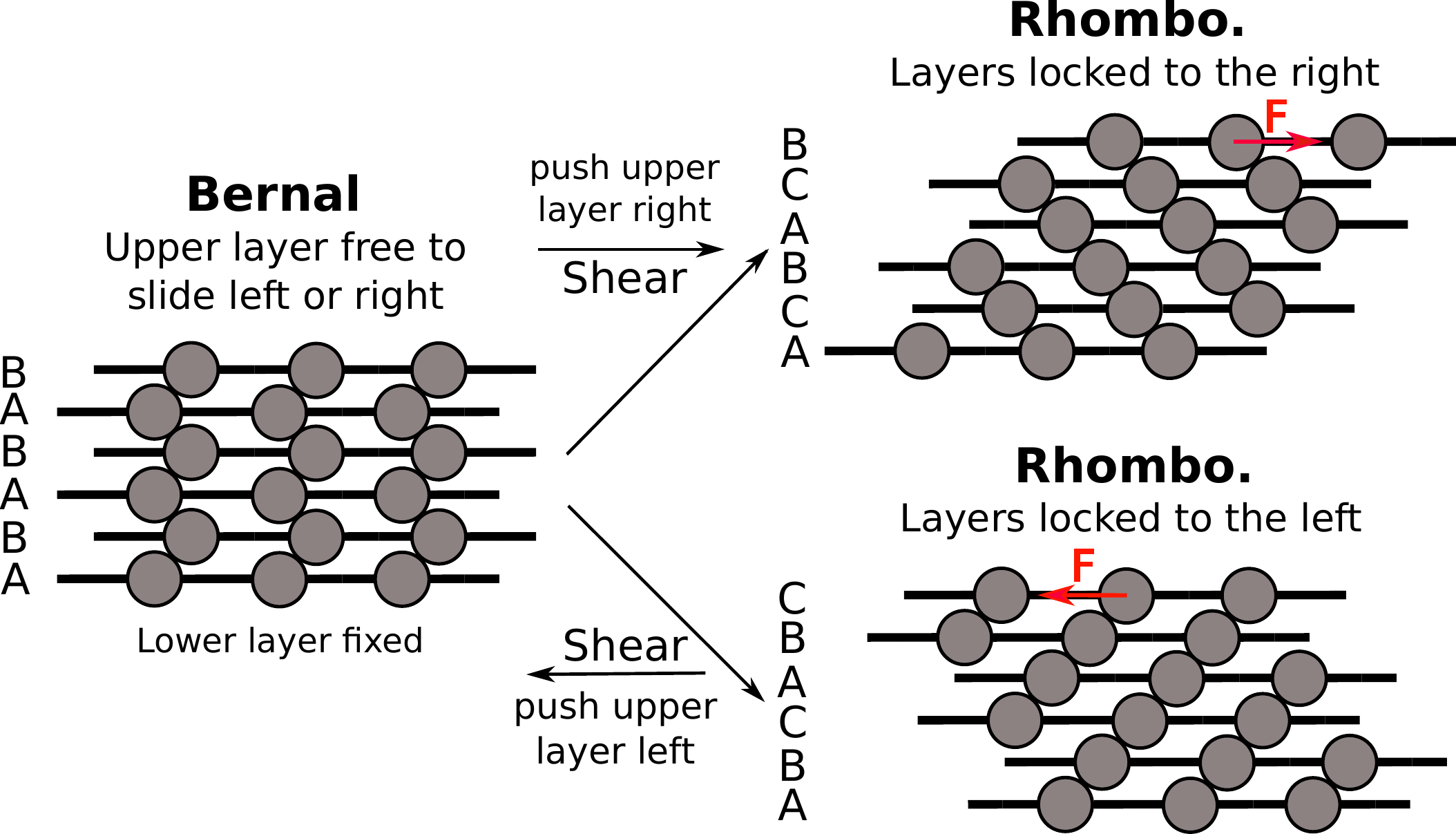}
\caption{\textbf{Transformation from BG to RG in mechanical model.} \textbf{a}, Black: One dimensional (1D) interaction energy per interface atom (see Methods), which we refer to as potential, of a two layer graphene system, where the upper layer moves with respect to the lower one along the armchair direction. 
Energies at $m_1$=AB and $m_2$=AC are the same and have the lowest energy. AA stacking (M) is the least favorable configuration, with energy $V_{AA}$. $b_1$=SP is a 
 barrier, with energy $V_{SP}$, that the upper layer needs to overcome to go from one minima to the other. Blue: Simplified square version of the black potential, where the lower barrier $V_{SP}$ is set to 0. The height going to infinity (hard wall) corresponds to the mechanical model in b and c. \textbf{b}, 1D mechanical model. The dark circles can be thought of as hard carbon atoms. Letters on the left label the position of each layer.
On the left, the upper layer can move without resistance to the right. After the upper layer is pushed to the right, it gets ``locked''. We refer to this as a sliding step.
 \textbf{c}, Analogous to b, but considering multiple layers.
The lower layer is fixed in position A.
The top layer is pushed to the right (upper right diagram). After successive sliding steps, all layers end up locked in the ACBACB configuration. Similarly, when pushing the upper layer to the left, layers end up locked in ABCABC.}
\label{fig:potential_1d}
\label{fig:mechanical_model}
\end{figure*}

\vspace{0.4cm}
\ni \textbf{\large Mechanical model}

\ni A simple mechanical model is used to explain the underlying mechanism in the transformation of multilayer BG to RG via shear stress. We start by considering the interaction energy of two layers of graphene, in which the upper layer moves relative to the lower one, fixed in what is referred to as position A. Calculations are carried out within density functional theory (DFT)  \cite{Charlier1994a,Lebedeva2011,Mounet2005,Zhou2015,Savini2011} with an LDA functional, since parameters like the shear frequency agree well with experiment (see Methods for details on why LDA is a good functional for our purposes). A layer of graphene on top of another one (configuration AA) corresponds to a maximum of interaction energy.
The most stable configuration is obtained when one of the graphene atoms of the upper layer is right above an atom of the lower layer, and the other atom is equidistant from six carbon atoms in the lower layer. There are two of these configurations, AB and AC, both corresponding to a Bernal bilayer. See Fig.~\ref{fig:mechanical_model}a. In configuration SP, the upper layer is in the middle of positions B and C. It corresponds to a saddle point in the full 2D bilayer energy, which we refer to as potential $V$. Throughout the whole paper, we consider the energy per interface atom (see Methods). The full energy curve, when moving the upper layer relative to the lower layer along the bond direction (also known as the armchair direction), starting from AA, results in the black curve of Fig.~\ref{fig:mechanical_model}a.

The mechanical model corresponds to
 a simplified version of the 1D potential of Fig.~\ref{fig:potential_1d}a: the low barrier around SP separating the two minima is neglected (flat region of width $2d$), while the high barrier around AA is considered as a hard wall (infinite potential of width $d$). It corresponds to the square potential in Fig.~\ref{fig:mechanical_model}a when the height goes to infinity. Then each layer, considering the interaction with an upper and lower layer, can be  considered just as a rigid block of width $d$ connected by rods of size $2d$. To make the visualization easier, we consider circles instead of blocks, and assume that each layer can only move horizontally. The circles can be thought of as hard carbon atoms. In Fig.\ref{fig:mechanical_model}b, the upper layer is free to move to the right, until it makes contact with the lower layer, getting ``locked''. This happens repeatedly when considering several layers, and translates into long-range rhombohedral order. If shear is applied as in the top part of Fig.~\ref{fig:mechanical_model}c, by exerting a force on the upper layer, ABABAB transforms into ACBACB (if the force is applied in the opposite direction, it transforms into ABCABC). That is, BG transforms into RG (a more detailed description is included in Methods).

\vspace{0.4cm}
\ni \textbf{\large Transformation from BG to RG: First-principles calculations}
\label{sec:transformation}

\begin{figure*}[!ht]
\centering
	{\includegraphics[width=0.65\textwidth]{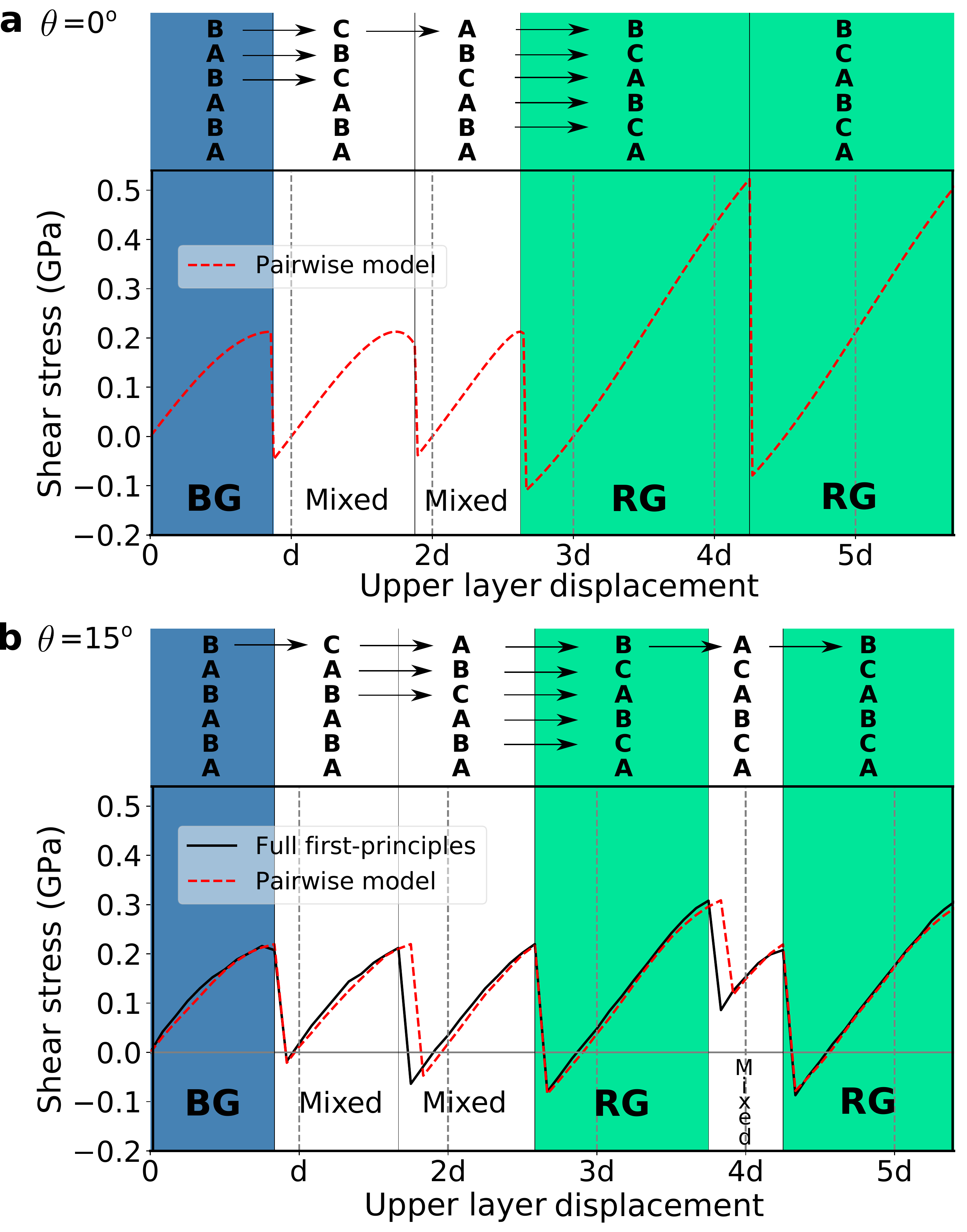}}
\caption{\textbf{Transformation from BG to RG in first principles calculation.} \textbf{a}, Shear stress along the armchair direction $y$ ($\theta=\ang{0}$) of a 6 layer calculation, as a function of the center of mass displacement of the upper layer. Only the interaction between nearest layers is considered. BG is transformed into RG by applying shear stress on the upper layer. The lower layer is fixed and the upper layer is moved and fixed in steps of $d/12$ (with $d$ the bond length) in $y$ (the perpendicular direction is relaxed). All other coordinates are relaxed. The letters on the top of each region indicate the stacking sequence into which the system relaxes to when the external stress is removed. The arrows indicate which layers change position. The initial configuration all in blue is BG, while the configuration all in green is RG.
As the upper layer moves in $+y$, it is pushed in $-y$ towards the original equilibrium configuration BG. The external stress increases until a sliding step takes place at the critical stress of about 0.2 GPa, and the stress decreases abruptly or ``jumps''.
After 3 sliding steps, RG is formed and layers are ``locked" in $+y$. Now the stress increases until about 0.5 GPa (big critical stress), the upper layer moves in the perpendicular direction $x$ (see Supplementary Fig.~\ref{fig:coord_6l}a), but the structure remains fully rhombohedral.
\textbf{b} Analogous to \textbf{a}, but with $\theta=\ang{15}$. The black curve is a full first-principles calculation. The excellent agreement between this pairwise model and the full first-principles calculation shows the pairwise model is sufficient to study transformations when shear is applied. RG is also formed after 3 sliding steps. Then, stress increases to about 0.3 GPa, the upper layer jumps in $x$, but now the system is not fully rhombohedral. After a sliding step however, RG is recovered, and the sequence continues repeating itself. Notice how the small critical stress does not depend much on the angle, whereas the big critical stress is significantly smaller for $\theta=\ang{15}$.
}
\label{fig:force_6l}
\end{figure*}

\ni Here we consider an analogous transformation to that of the mechanical model of the previous section, using  two calculations. In one case, we consider only the interaction between nearest layers, in what we refer to as the pairwise model. The other is a full DFT-LDA calculation. The calculations agree very well (see Fig.~\ref{fig:force_6l}b), showing that the pairwise potential is sufficient to study how the layering sequence changes with shear stress. The main qualitative difference  with the one dimensional mechanical model of the previous section is that, after layers are locked in RG, they move in the perpendicular direction if the external stress increases too much.

\begin{figure*}
\centering
	{\includegraphics[width=1\textwidth]{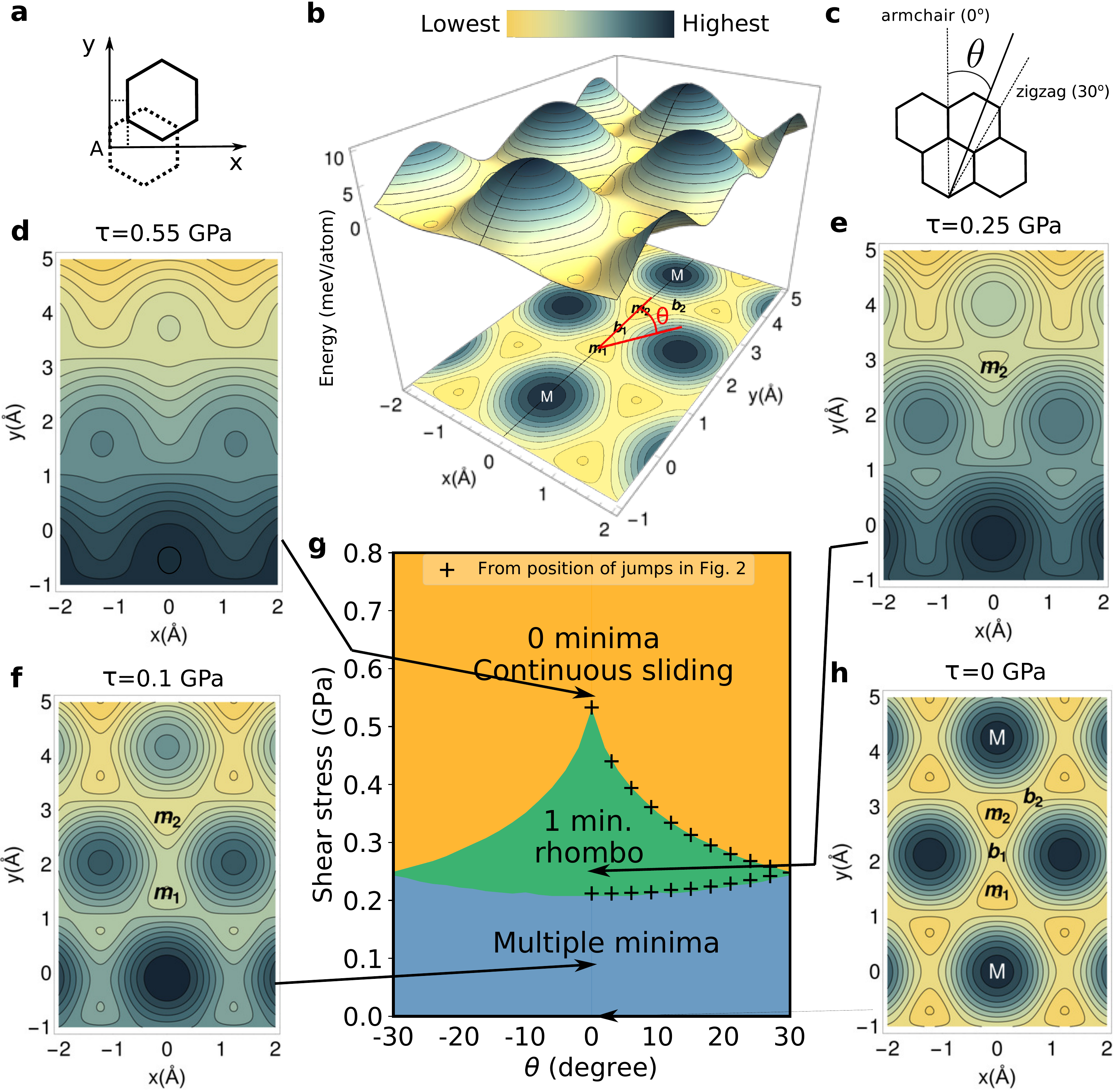}}
\caption{\textbf{Bilayer potential phase diagram}. \textbf{b}, Energy per interface atom $V$ (DFT fit) of a bilayer graphene system, with the center of mass of the upper layer moving relative to the lower layer. Energy is higher in darker regions and lower in lighter regions, as indicated in the color bar. The lower layer is fixed in position A, and \textbf{a} indicates the coordinate system. A projection on the plane is also displayed below. The slice $x=0$ is shown in black, just as in Fig.~\ref{fig:potential_1d}a. \textbf{c}, The angle $\theta$ indicates the direction of the shear stress $\pmb{\tau}$ with respect to the armchair direction.  \textbf{g}, Stress-angle phase diagram obtained by counting the number of minima of the enthalpy $H(\mathbf{r})= V(\mathbf{r}) - \pmb{\tau} \cdot \mathbf{r} A$, with $A$ the area of the flake. (i) Blue region: 2 minima (multiple minima in a $N$ layer system). System remains in the current local minima, which is BG if that is the starting point (the most stable structure when there is no stress). (ii) Green region: 1 minimum. The relative position of a layer relative to the lower one is always the same, so the phase is RG. (iii) Orange region: There is no local minima, so the system slides continuously. The plus signs `+' are obtained from calculations as in Fig. 2, and correspond to the critical values of stress before it decreases abruptly. For each angle, the small critical stress gives the lower plus sign, and the big critical stress the upper one. We see they agree very well with the borders between the regions determined from the minima analysis (lower crosses match the blue-green border, and upper crosses the green-orange border). Thus, the minima analysis is sufficient to characterize the system. A transition to RG may occur at lower values due to thermal fluctuations. \textbf{h}, Same as contour plot in \textbf{b}. There are two equivalent minima $m_1$ and $m_2$, separated by barriers $b_1$ and $b_2$. \textbf{f}, \textbf{e} and \textbf{d} show $H$ as a function of $x$ and $y$ for values of stress corresponding to the three regions, with 2, 1 and 0 minima, respectively.
}
\label{fig:potential_2d}
\label{fig:phase_diagram}
\end{figure*}

The center of mass of the lower layer is fixed in all calculations (it can be thought of as attached to a substrate like copper or nickel, that have a larger shear stress), 
 and the upper layer is ``pushed'' 
 by a fraction of the bond length $d$ along the direction that makes an angle $\theta$ with the armchair direction  (see Fig.~\ref{fig:phase_diagram}a). In each step, the structure is relaxed (more details in Methods).
Fig.~\ref{fig:force_6l} shows the external force per unit area  (shear stress) on the center of mass of the upper layer, for $\theta=\ang{0}$ (a) and $\theta=\ang{15}$ (b) (the component of the force perpendicular to the $\theta$ component is 0, since the upper layer is relaxed in that direction). We consider a quasistatic transformation, so the external force is minus the force exerted by the rest of the system.
 
The initial configuration is ABABAB (six layers of BG, in blue). Let us first consider $\theta=\ang{0}$.
As the upper layer starts to move in $+y$ from its initial position B, the rest of the system 
 tries to restore it to the equilibrium position BG, and stress increases. When it reaches the critical stress of about 0.2 GPa (which we refer to as small critical stress), it drops abruptly, and the system moves towards the nearest minimum, ABABAC. A sliding step has taken place, analogous to Fig.~\ref{fig:mechanical_model}b. In the literature, this type of gradual movement followed by sudden jumps (see also Supplementary Fig. 1) is known as ``stick-slip'' motion \cite{Dienwiebel2004, Verhoeven2004}.
The regions are delimited by the points where the force drops abruptly, and are labeled on top by the structure the system relaxes to upon releasing the external force. The arrows indicate the layers that are changing position from one region to the next one. After the first sliding step, two additional sliding steps take place, and RG (green) is formed, so all layers are locked. The stress increases now until a larger critical stress of around 0.5 GPa (big critical stress), the upper layer suddenly changes $x_\textrm{CM}$ (Supplementary Fig.~\ref{fig:coord_6l}a), moving ``around'' the maxima that the high-stress configuration is close to, and the force decreases significantly. In this case, rhombohedral order remains. For $\theta=\ang{15}$, after 3 sliding steps RG is also formed, but after the big critical stress rhombohedral order is partially lost. However, after one sliding step RG is obtained again.
The small critical stress is around 0.2 GPa, similar to that of $\theta=\ang{0}$, but the big critical stress is around 0.3 GPa, which differs considerably from 0.5 GPa. There is indeed a significant angle dependence for the big critical stress, as can be observed later in more detail in Fig.~\ref{fig:phase_diagram}g.

Thus, if the magnitude of the applied stress is lower than the small critical stress, around 0.2 GPa, the layering will not change. It will stay as BG after removing the stress. If stress is between the small and big critical stresses, the final structure will be RG. If the applied stress is larger than these limiting upper values
, layers will keep on sliding.

\vspace{0.4cm}
\ni \textbf{\large Full 2D bilayer potential}

\ni The excellent agreement between the pairwise model and the first-principles calculations (Fig.~\ref{fig:force_6l}b) suggests it should be possible to characterize a $N$ layer system in terms of the building block of the pairwise model, the potential $V$, shown in Fig.~\ref{fig:potential_2d}b. We will now show this is indeed the case. $V$ was obtained by considering the upper layer in different positions with respect to the lower one (see details in Methods). The potential along the $x=0$ line, shown in black, is the same as in Fig.~\ref{fig:mechanical_model}a. Fig.~\ref{fig:potential_2d}a indicates the system of coordinates: the lower layer (dashed) is fixed and determines the origin, while the center of mass position of the upper layer determines the $x,y$ coordinates.

\vspace{0.3cm}
\ni \textbf{Phase diagram.} As mentioned earlier, depending on the magnitude and angle of the stress applied, the system can be BG, RG, or slides continuously. When the system is subjected to a shear stress $\pmb{\tau}=\mathbf{F}/A$, where $\mathbf{F}$ is the applied force and $A$ the area of the flake, it can be studied using the bilayer enthalpy $H(\mathbf{r}) = V(\mathbf{r}) - \pmb{\tau} \cdot \mathbf{r} A$. As we will now see, the number of minima of $H$ determines the phase the system is in, giving a stress-angle phase diagram (Fig.~\ref{fig:potential_2d}g) for multilayer graphene.

In the pairwise model, if a stress $\pmb{\tau}$ is applied to the upper layer in a quasistatic transformation, then the layer below exerts all the remaining stress $-\pmb{\tau}$. The same applies to subsequent layers. Thus, for each pair of layers, their enthalpy $H$ is the same. Depending on the angle and magnitude of $\pmb{\tau}$, there are three possible situations, which correspond to the 3 colored regions of Fig.~\ref{fig:potential_2d}g:\\
(i) Blue region: 2 minima. If no stress is applied, $H=V$, and there are 2 minima $m_1$ and $m_2$ (see Fig.~\ref{fig:potential_2d}h). If shear stress is sufficiently low, $H$ still has 2 minima, and the system has $2^{N-1}$ minima. For each pair of layers, they do not escape the local minimum they are currently in. Since this holds for all layers, the full system does not escape the local minimum it is currently in. In particular, if BG is chosen as the starting structure, the system remains BG in the blue region (in the pairwise model, all stacking sequences have the same energy, but in reality BG is the most stable structure).\\
(ii) Green region: 1 minimum. As stress increases, $m_1$ disappears and $m_2$ remains close to its $\pmb{\tau}=0$ position. Barrier $b_1$ disappears, while $b_2$ remains. This occurs because the stress is more aligned with the direction in which the saddle point $b_1$ has a maximum (that is, the direction in which $b_1$ acts as a barrier) than with the corresponding direction of $b_2$ (except at $\theta=\ang{30}$, where both barriers are affected in the same way, and the system transitions directly from 2 minima to 0 minima). Since there is only one minimum, all layers are in the same position relative to the lower layer, and the resulting stacking is rhombohedral. This is a key observation of our work. From the convention in Fig.~\ref{fig:mechanical_model}a, $m_2$ corresponds to configuration AC, and layers are ACB-stacked. If stress is applied in the opposite direction, $m_1$ is the only remaining minimum, and layers are ABC-stacked.\\
(iii) Orange region: 0 minima. For larger stresses, there are 0 minima. Since there are no local minima, layers keep on sliding without reaching a stable configuration.
If the stress is eventually removed, the system will not necessarily be RG.
However, transformations at different angles as in Fig.~\ref{fig:force_6l} suggest that the system will still have a high degree of rhombohedral order.

In particular, for small angles, the structure remains fully RG. Also, $\theta=\ang{0}$ is the angle with the largest range of stress that results in RG, of about 0.3 GPa (from 0.2 to 0.5 GPa). Thus, the armchair direction is the most robust direction to obtain RG.

Fig.~\ref{fig:phase_diagram}g also shows with plus signs '+' the critical stress values. For each angle, the small critical stress corresponds to the lower value, and the big critical stress to the upper value. They coincide with the blue-green border, and green-orange border, respectively. This excellent agreement shows that the number of minima of $H$ does indeed define the stress-angle phase diagram.

It is worth pointing out that when $m_1$ becomes shallow enough, it might be possible for thermal fluctuations to excite layers from $m_1$ from $m_2$. This will depend on experimental conditions, like duration of the experiment, temperature, and size of the flakes. Thus, the curve that separates multiple minima from 1 minimum is actually an upper bound. Also, BG is more stable than RG and is presumably located in a deeper (local) minima (in an analogous fashion to the two minima of the blue curve of Supplementary Fig.~\ref{fig:6layer}). So for $\theta$ close to 30$^\circ$, it might occur that the system transitions directly from BG to continuous sliding.

\vspace{0.3cm}
\ni \textbf{Pressure.} Other hydrostatic pressures were also considered.
Supplementary Fig.~\ref{fig:force_pressure} shows the phase boundaries at P=0 (blue-green and green-orange borders of Fig.~\ref{fig:phase_diagram}g, or lower and upper borders) and P=2 GPa. The values of the lower and upper boundary at $\theta=0$
increase approximately linearly with pressure, at about 0.07 and 0.18 GPa of shear stress per GPa of hydrostatic pressure, respectively. Thus, the amount of stress needed to obtain RG increases, but also the range of allowed values to obtain RG (which might increase the robustness of an experiment).


\vspace{0.3cm}
\ni \textbf{Shear in previous works and superlubricity.} The values of stress to produce RG suggested by our calculations  are very similar to values already published in experimental reports.
The configurations we have described,
where layers are commensurate with each other, are referred to as \textit{lock-in} states in graphene literature related to friction or shear. In this type of systems, values of shear stress of the order of 0.1 GPa were measured \cite{Dienwiebel2004}, and observed to be in good agreement with previous calculations\cite{Verhoeven2004}.
In another experiment\cite{Liu2012b}, a microtip of a micromanipulator was used to apply a shear force on a graphene flake to ``unlock'' it (remove it from the minima), and based on the deformation of the tip, a value of 0.14 GPa was reported, also lower than 0.20 GPa .
On the other hand, when the layers are incommensurate with each other, the values of friction are 2 or 3 orders of magnitude lower. This phenomenon is referred to as superlubricity and has sparked a lot of interest.
Optimal conditions for superlubricity include big flakes, low temperatures and low loads\cite{Wijn2010}. Since layers have to be moved out of the local minimum in the mechanism we have proposed, depending on the experimental conditions, care might need to be taken to avoid the upper layer to rotate into a superlubricant state.

To conclude, we have described a mechanism to transform multilayer graphene into RG through shear stress, which implies that applying sufficient shear strain to graphite results in long-range RG. Also, existing experimental values of shear stress are similar to the ones suggested by our results.
Our model suggests a compelling method
for experimental groups trying to obtain multiple layers of rhombohedral-stacked graphene.

\clearpage
\small
\ni \textbf{\large Methods}

\ni \textbf{DFT calculations.}
Calculations were performed in Quantum Espresso (QE) \cite{QE} using an LDA functional. The bilayer potential and the six layer first-principles transformations were obtained with a cutoff of 80 Ry, a $k$-grid of $56\times 56\times 1$ and an electronic temperature (Fermi-Dirac smearing) of 284 K. The energies considered in this work are always in meV per interface atom. This means that the total energy of the system is divided by 2, whether the number of layers is 2 or 6 (which permits a direct comparison between energies of both systems). In each system, the energy is set to 0 at the relative position where the energy is the lowest. To converge the energy differences of Supplementary Table 1 with a precision below 0.01 meV/atom, we used an energy cutoff of 120 Ry and an electronic grid of $80 \times 80 \times 1$. In the rvv10 calculations of Supplementary Fig.~\ref{fig:rvv10}, 100 Ry were used.

The most important parameter in the calculations is basically the small barrier $V_\textrm{SP}$ of Fig.~\ref{fig:mechanical_model}a between the two minima, since it determines the amount of stress necessary to go from one local minimum to another. The value of $V_\textrm{SP}$ can be expected to be somewhat accurate if the curvature around the minima is accurate. The curvature is directly proportional to the frequency squared of the shear mode LO' at $\Gamma$ \cite{Mounet2005} (this is the mode in which layers move in the plane in opposite directions). Experimental values of the shear frequency vary between 42 cm$^{-1}$ and 45 cm$^{-1}$(ref. \cite{Lebedeva2011}). This is in good agreement with the value that results from fitting a parabola close to $m_1$ in Fig.~\ref{fig:potential_1d}a, 42 cm$^{-1}$. The curvature of other functionals like rvv10 (ref. \cite{Sabatini2013}), which includes van der Waals, is actually lower in our calculations, so they agree less well with experiments (see Supplementary Fig.~\ref{fig:rvv10}). In previous works, rvv10 also gives a lower frequency than LDA\cite{Grande2019}.

To be confident that it is sufficient to consider the 2 layer potential in our analysis, we considered a similar calculation, but with 6 layers (dotted-blue line in Supplementary Fig.~\ref{fig:potential_1d}a) instead of 2. In each calculation, the $x_\textrm{CM}$ and $y_\textrm{CM}$ of each of the three upper layers was moved with respect to 
the three lower layers, and the system was relaxed. This corresponds to a generalized stacking fault energy \cite{Wang2015}. The curves are very similar, and the energy difference between the minima is about 6\% of $V_\textrm{SP}$, so the transformations could be analyzed in terms of the bilayer potential. Indeed the curves in Fig.~\ref{fig:force_6l} are almost identical. Also, the value obtained for the stacking fault energy, 1.58 meV/atom, compares well with 1.53 meV/atom obtained with RPA \cite{Zhou2015}. A detailed comparison is included in Supplementary Table 1.

\vspace{0.3cm}
\ni \textbf{Mechanical model.} Let us label L$i$ the $i$th layer from the bottom to the top, and let us consider that a force is applied to the right on the upper layer L6, as in the top part of Fig.~\ref{fig:mechanical_model}c.
First, L6 moves from B to C without resistance, where it locks with L5. Then, it pushes L5 from A to B, which in turn pushes L4 from B to C. The lower layers have not moved yet. In the next step, L2 and L3 move as well. The transformation is (labeling always from bottom to top): ABABAB$\rightarrow$ABABAC$\rightarrow$ABACBA$\rightarrow$ACBACB. If the direction is reversed, the steps of the transformation are: ABABAB$\rightarrow$ABABCA$\rightarrow$ABCABC. In both cases, RG is obtained.

\vspace{0.3cm}
\ni \textbf{Transformation calculations.}
In Fig.~\ref{fig:force_6l}, in addition to moving the upper layer in steps of $d/12$ along the direction $\theta$ (and fixing it, while relaxing in the perpendicular direction), where $d$ is the bond length, after each step layer $i=2,..,5$ is actually moved as well in $[(i-1)/5]d/12$ (other layers are fully relaxed), giving a configuration closer to the equilibrium position. This gives a more accurate value of the position at which the abrupt changes in the stress take place, and which particular layers slide relative to each other in $y$ or in the perpendicular direction $x$. 

At $\theta=\ang{0}$ there is a bifurcation of behavior when the big jump occurs (with the upper layer jumping in $-a/2$ for $\theta <0$, and $a/2$ for $\theta>0$). So the result of Fig.~\ref{fig:force_6l}a actually corresponds to $\theta= \ang{0.1}$, to avoid an artificial behavior at $\theta=\ang{0}$.

\vspace{0.3cm}
\ni \textbf{Fourier interpolation.}
\label{sec:Fourier}
To obtain the bilayer potential at an arbitrary $\mathbf{r}$ point as in Fig.~\ref{fig:potential_2d}b (see coordinate system there), and to then obtain the phase diagram, the potential was first calculated at a set of $N$ discrete points $\mathbf{r}_i=(x_{\textrm{CM},i},y_{\textrm{CM},i})$ of the upper layer within a primitive cell (separated by $a$/15 along the lattice vectors, with $a=2.46$ \AA\hspace{1mm} the length of the lattice vectors, totaling $N=225$). As mentioned earlier, the interlayer distance and relative coordinates were relaxed. Let $\mathbf{G}$ be a set of $N$ reciprocal lattice vectors around the origin with the symmetry of the crystal. Then, we can write

\be
\begin{split}
V(\mathbf{r}) = \f{1}{N} \sum_\mathbf{G} V_\mathbf{G} e^{i \mathbf{G}.\mathbf{r}} \\
\textrm{with } V_\mathbf{G} = \sum_\mathbf{r_{i}} V(\mathbf{r_i}) e^{-i \mathbf{G}.\mathbf{r_i}} \\
\end{split}
\ee

\ni where $\mathbf{r}$ can take any value.

\vspace{0.3cm}
\ni \textbf{Hydrostatic pressure.}
In order to obtain a phase diagram as in Fig.~\ref{fig:phase_diagram}g when there is an external pressure $P$, we have to add an additional term to the enthalpy $H$ defined earlier, resulting in the new enthalpy

\be
H(\mathbf{r}) = V(\mathbf{r}) - \pmb{\tau} \cdot \mathbf{r} A + P z A,
\ee

\ni where $z$ is the interlayer distance of the center of mass at pressure $P$.

\vspace{0.3cm}
\ni \textbf{Acknowledgements}

\ni The authors would like to thank Stony Brook Research Computing and Cyberinfrastructure, and the Institute for Advanced Computational Science at Stony Brook University for access to the high-performance SeaWulf computing system, which was made possible by a \$1.4M National Science Foundation grant (\#1531492).
We also acknowledge support from the Graphene Flagship
Core 2 grant number 785219 and Agence Nationale de
la Recherche under references ANR-17-CE24-0030. This
work was performed using HPC resources from GENCI,
TGCC, CINES, IDRIS (Grant 2019-A0050901202) and PRACE.

\vspace{0.3cm}
\ni \textbf{Author Contributions}

\ni The project was conceived by all authors. JP.N. performed the numerical calculations, prepared the figures and wrote the paper with inputs from M.C. and F.M. All authors contributed to the editing of the manuscript.

\vspace{0.3cm}
\ni \textbf{Competing Interests}

\ni The authors declare no competing interests.


\setcounter{figure}{0} 

\begin{figure*}[!ht]
\begin{figCaption}
\centering
	{\includegraphics[width=0.5\textwidth]{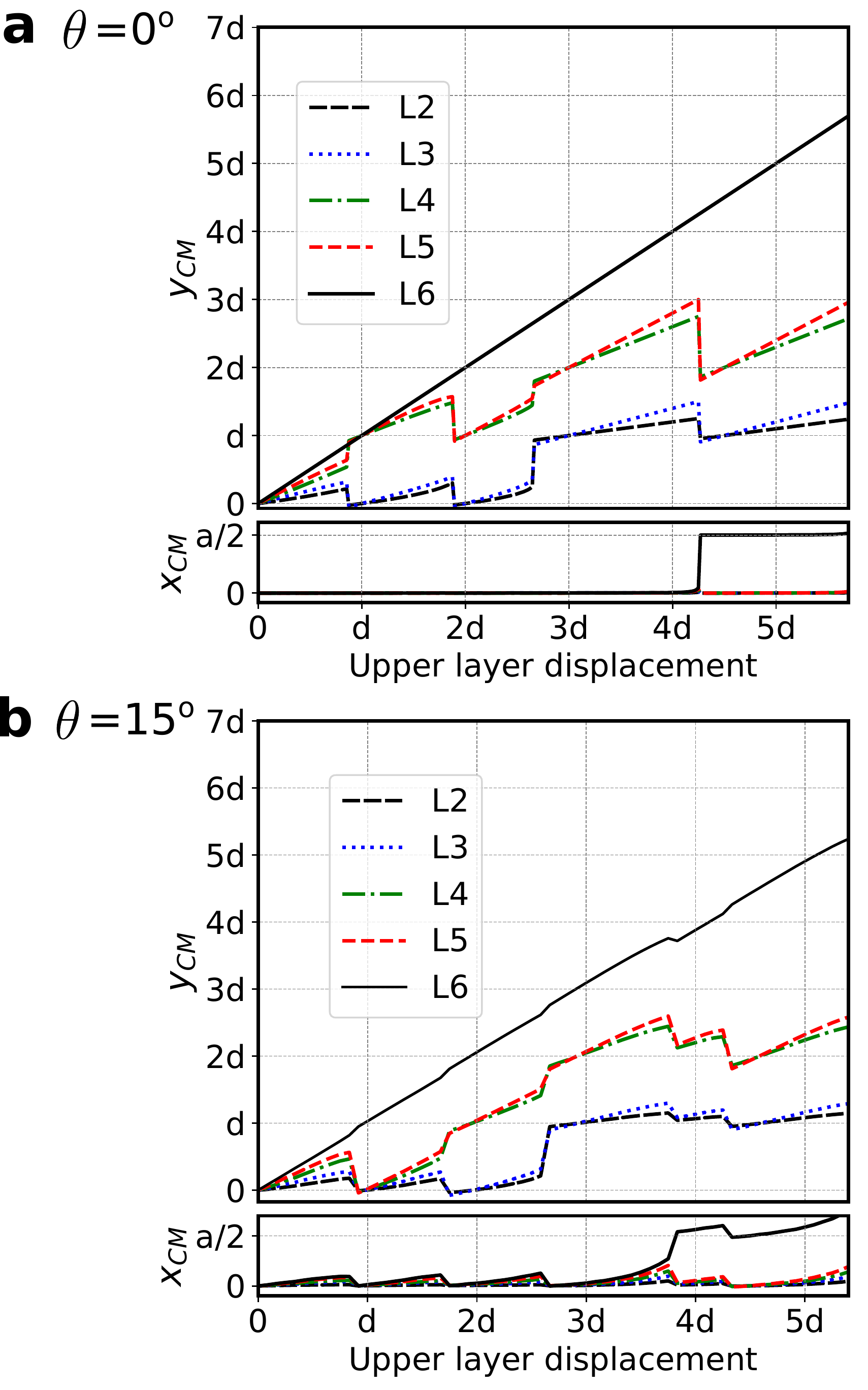}}
\caption{
\textbf{a}, Change of $y_\textrm{CM}$	and $x_\textrm{CM}$ of each layer, as a function of the center of mass position $y_\textrm{CM}$ of the upper layer (L6), for the 6 layer pairwise model calculation (see Fig.~\ref{fig:force_6l}a). Starting from ABABAB, the upper layer is ``pushed'' along the bond direction: 1/12 of the bond length $d$ is added to $y_{\textrm{CM}}$ of the upper layer in each step, and then the structure is relaxed. The relaxed coordinates are used in the subsequent calculation. In the first sliding steps, layers only move in $y$. After RG is formed, 
stress increases until the upper layers jumps in $x$ in $a/2$. \textbf{b}, Analogous plot, with $\theta=\ang{15}$. In this case, the structure does not remain fully rhombohedral after the jump in $x$, but is soon recovered after a sliding step in $y$.}
\label{fig:coord_6l}
\end{figCaption}
\end{figure*}

\begin{figure*}[h!]
\begin{figCaption}
\centering
	{\includegraphics[width=0.5\textwidth]{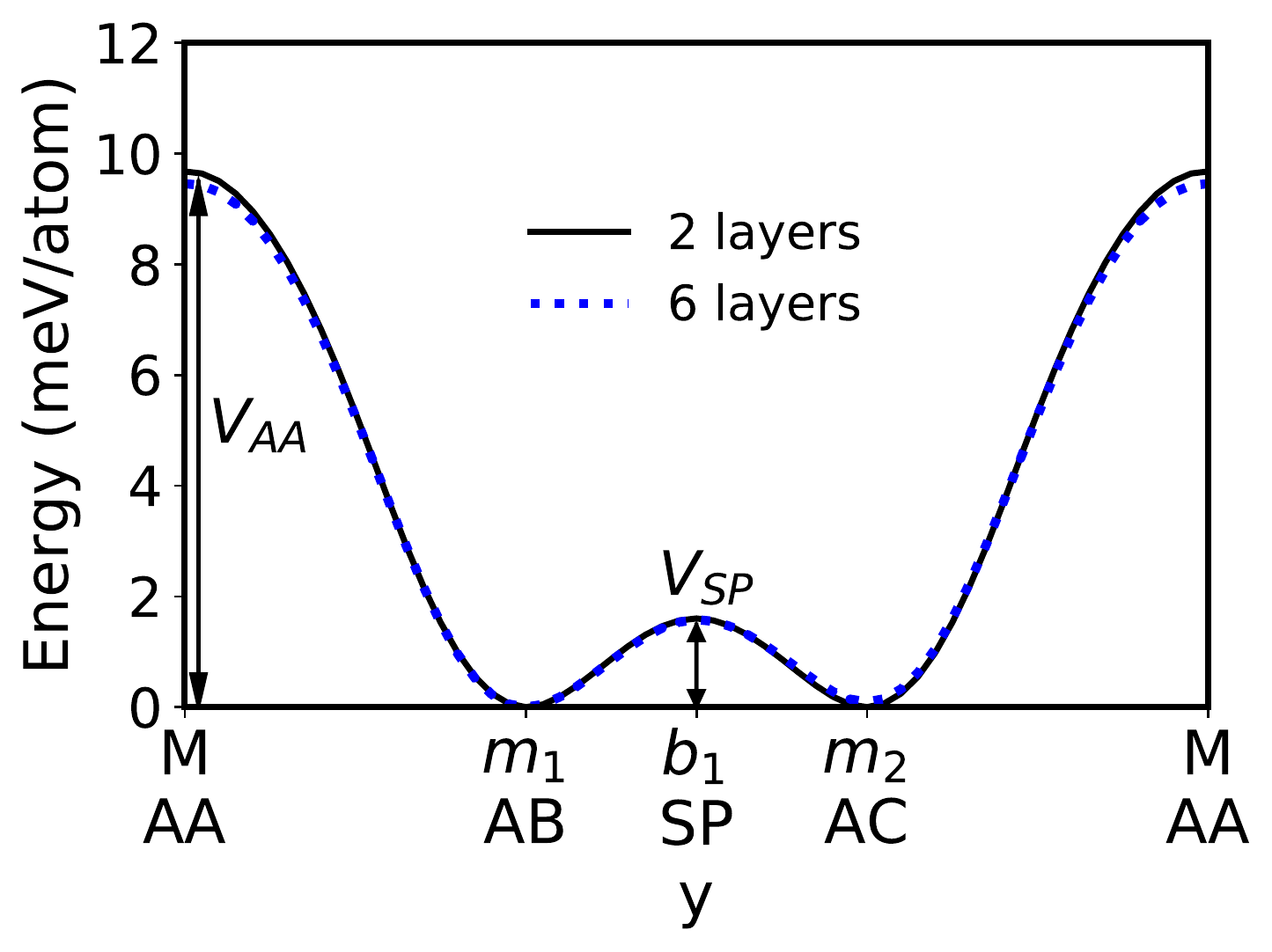}}
\caption{Full-black: Energy of the bilayer system along the armchair direction. Same as in Fig.~\ref{fig:mechanical_model}a.
Dotted-blue: Similar calculation with 6 as
opposed to 2 layers. The center of mass of the lower layers is fixed in the configuration ABA in all calculations, while the upper
3 layers move (the center of mass of each layer is fixed in each calculation, while other coordinates are relaxed). Labels $m_1$
and $m_2$ correspond to ABABAB and ABACBC, respectively. The energy difference corresponds to a stacking fault energy. M
corresponds to ABAACA. The similarity between the curves indicates that the 2 layer potential is a good approximation to
analyze how shear affects the stacking order. See Supplementary Table I for a more detailed comparison.}
\label{fig:6layer}
\end{figCaption}
\end{figure*}

\begin{figure*}[t]
\begin{figCaption}
\centering
	{\includegraphics[width=0.5\textwidth]{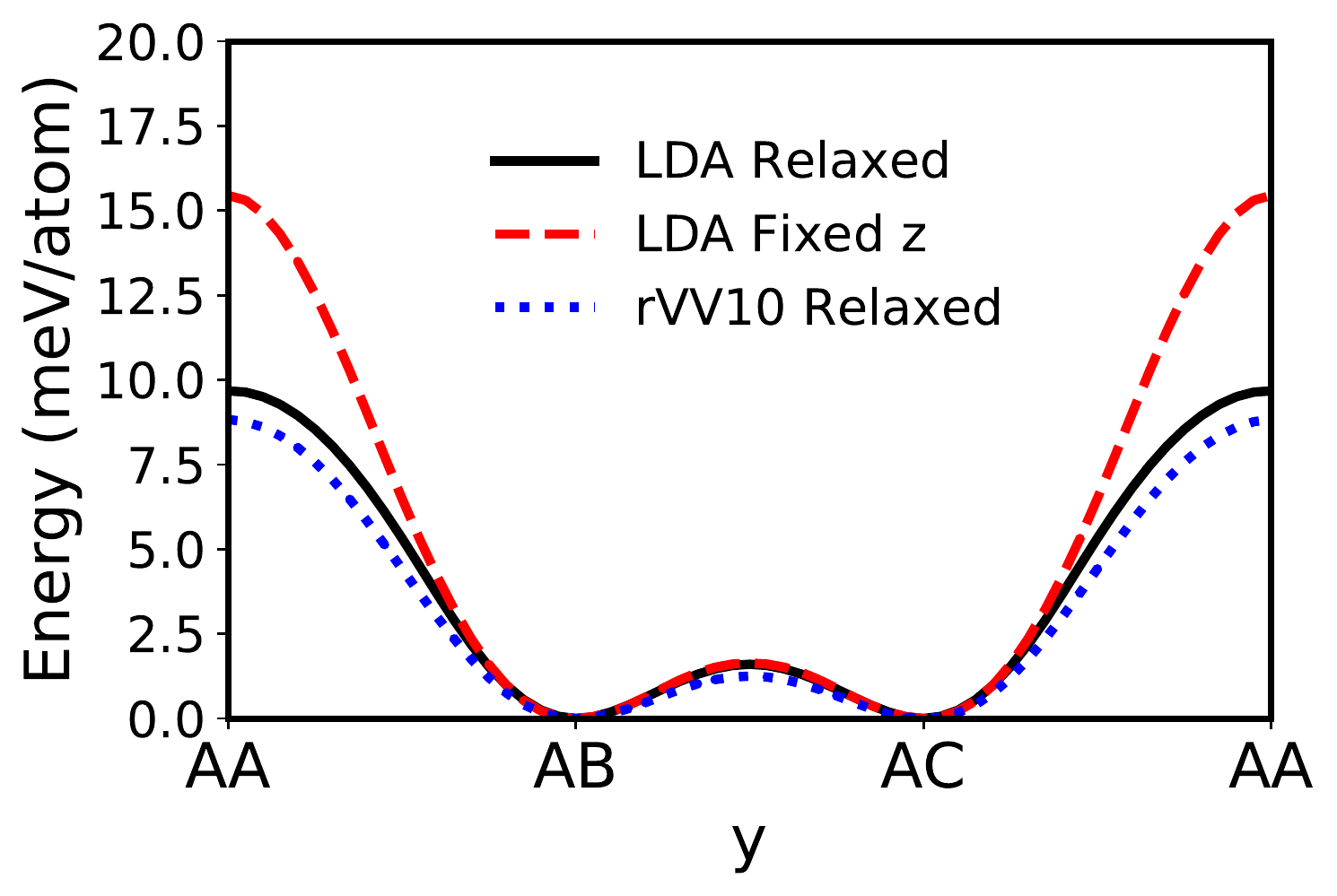}}
\caption{Energy of the bilayer system along the armchair direction. Full-black: LDA relaxed calculation (same as Fig.~\ref{fig:potential_1d}a). Dashed-red: Fixed z (at the relaxed value of the AB configuration). Dotted-blue: Relaxed rvv10. Although in the AB-AC region the agreement is good, the disagreement between both curves increases as the configuration approaches AA. This is consistent with Supplementary Fig.~\ref{fig:dist_pressure}, which shows the variation of the interlayer distance $z_\textrm{CM}$.
$z_\textrm{CM}$ at the small barrier is 3.33 \AA, while at AA  is 3.60 \AA. Thus, it is important to relax the configuration. Otherwise, the potential differences are artificially high.}
\label{fig:rvv10}
\end{figCaption}
\end{figure*}

\begin{figure*}[t]
\begin{figCaption}
\centering
	{\includegraphics[width=0.5\textwidth]{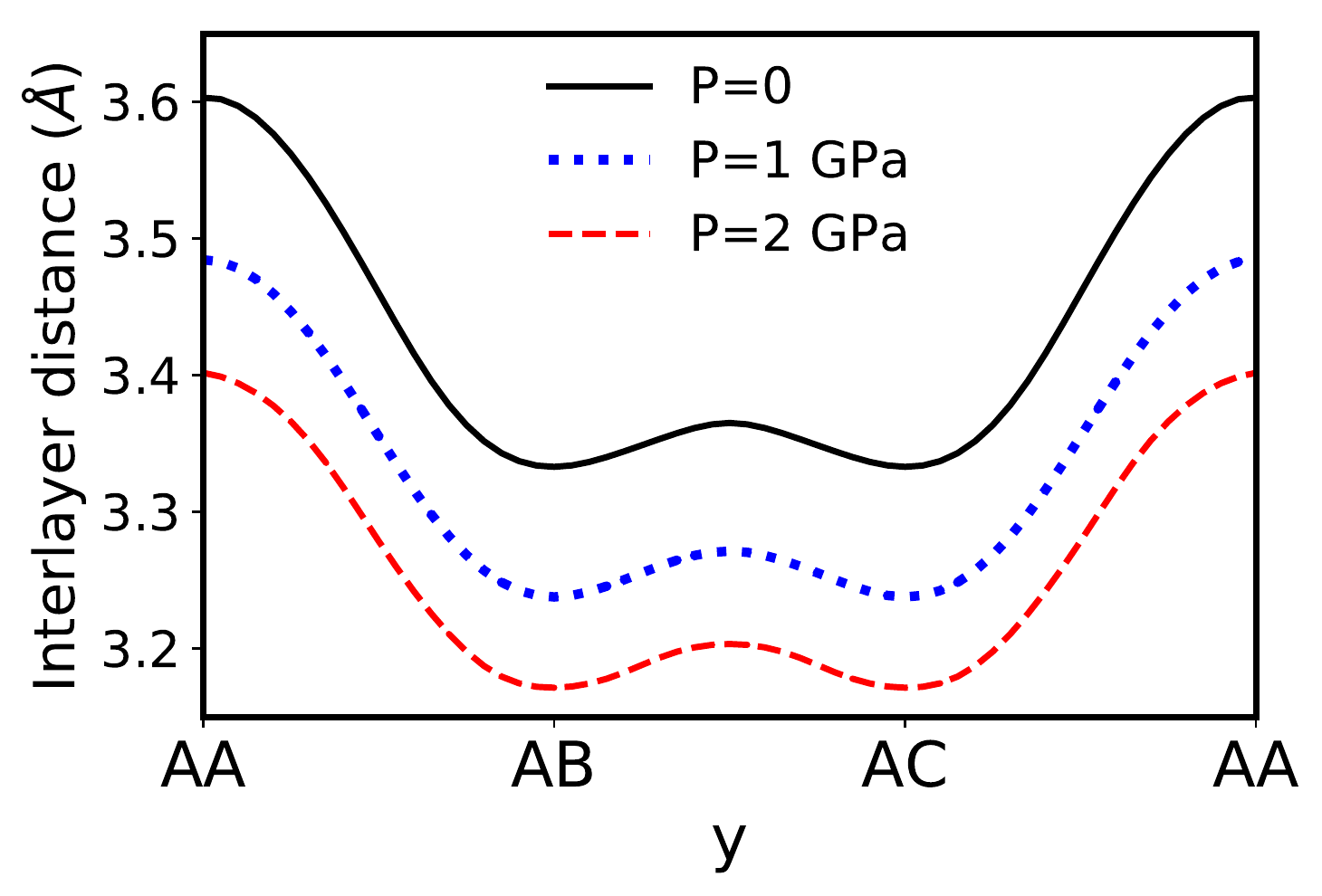}}
\caption{Interlayer distance of the bilayer system along the armchair direction, for various hydrostatic pressures. Full-black: $P=0$. Dotted-blue: $P=$ 1 GPa. Dashed-red: $P=$ 2 GPa. }
\label{fig:dist_pressure}
\end{figCaption}
\end{figure*}

\begin{figure*}
\begin{figCaption}
\centering
	{\includegraphics[width=0.5\textwidth]{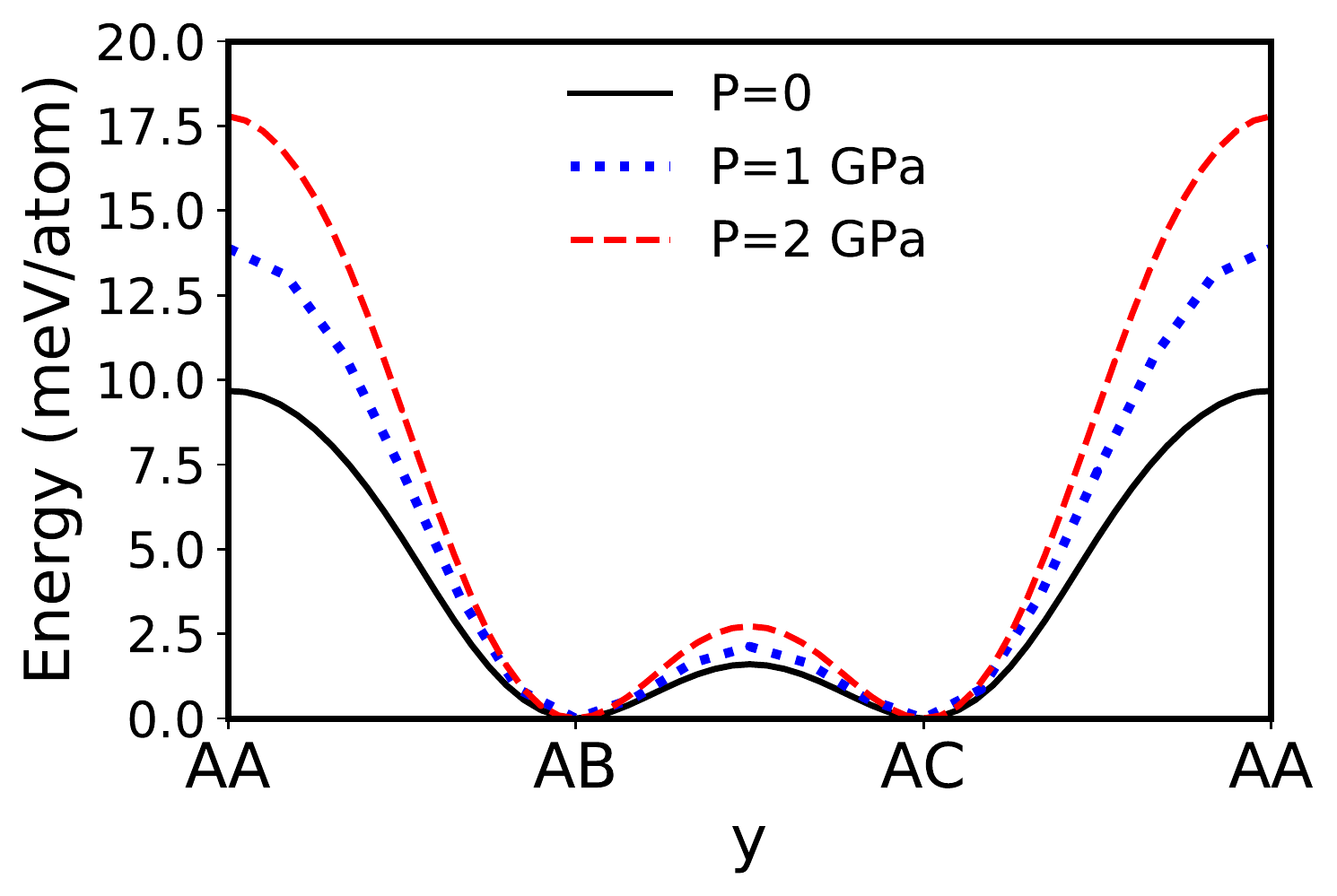}}
\caption{Plots analogous to those of Fig.~\ref{fig:mechanical_model}a at varying pressures. Full-black: $P=0$ (same as Fig.~\ref{fig:potential_1d}a). Dotted-blue: $P=$ 1 GPa. Dashed-red: $P=$ 2 GPa. Supplementary Fig. 3 shows that fixed $z$ close to AA corresponds to artificial pressures of over 1 GPa.}
\label{fig:energy_pressure}
\end{figCaption}
\end{figure*}

\begin{figure*}
\begin{figCaption}
\centering
	{\includegraphics[width=0.5\textwidth]{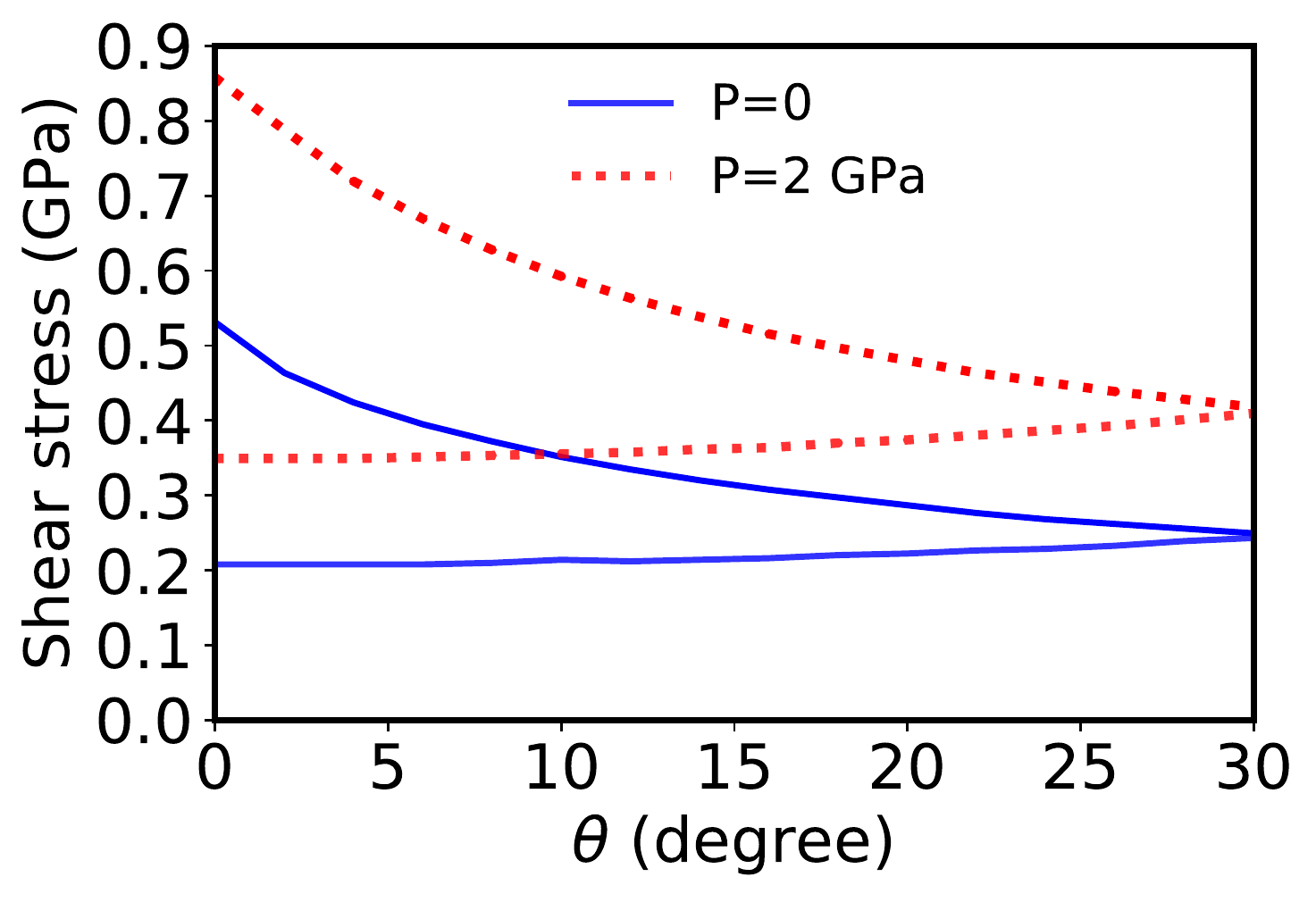}}
\caption{Phase boundaries at different hydrostatic pressures. For each color, the lower line corresponds to the border between the blue and green region of the phase diagram (Fig.~\ref{fig:phase_diagram}g), and the upper curve to the border between the green and orange regions. Blue: $P=0$ (same as in Fig.~\ref{fig:phase_diagram}f). Red: $P$= 2 GPa. In this range of pressures, the values of the curves at $\theta=\ang{0}$ increase linearly with pressure.
}
\label{fig:force_pressure}
\end{figCaption}
\end{figure*}

\begin{figure*}[!ht]
\begin{figCaption}
\centering
	{\includegraphics[width=0.5\textwidth]{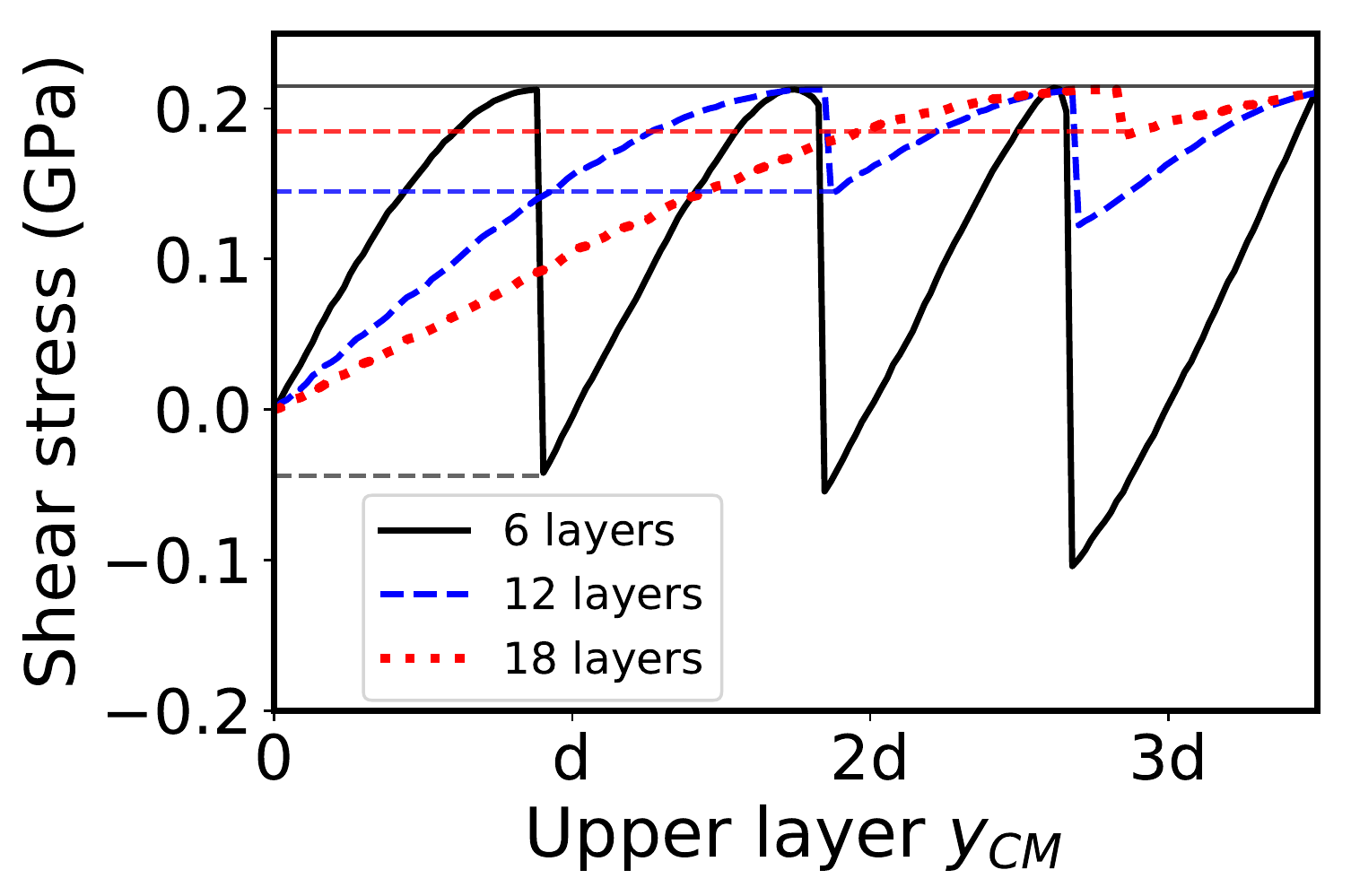}}
\caption{Shear stress on the upper layer as a function of $y_\textrm{CM}$ of the upper layer. The jump of stress, as sliding steps take place, decreases with the number of layers. The jump is also smaller after RG is formed.
This can be useful to apply a less varying and more controllable stress. If for example a cantilever is used to move the upper layer, the number of layers of the sample would have to be taken into account to decide on its optimal elastic constant. It can also be seen how the position of the first jump increases linearly with the number of layers.}
\end{figCaption}
\label{fig:jump}
\end{figure*}

\newpage

\begin{table*}[h]
\begin{tabCaption}
\caption{Comparison between the 2 layer potential and the 6 layer calculation of Fig.~\ref{fig:potential_1d}, and previous works. Ref.~\cite{Wang2015,Zhou2015} use the adiabatic-connection fluctuation-dissipation theorem within the random phase approximation (ACFDT-RPA). Ref.~\cite{Telling2003} uses anisotropic elasticity theory, revised values for the elastic constants and the experimental data of ref.~\cite{Baker1961,Amelinckx1965}, to obtain an average stacking fault energy of
0.14 meV per interface atom, the same value as the RPA calculation. The difference between the minima (stacking fault energy) is small compared to the barrier $V_\textrm{SP}$ separating them, which implies the 2 layer potential provides a good approximation to study the transformation through shear.}
{\begin{tabular}{cccc}
\hline \hline
\multicolumn{1}{l}{}  & \multicolumn{3}{c}{Energy (meV/atom)} \\
\multicolumn{1}{l}{}  & \begin{tabular}[c]{@{}c@{}}2 layer\\ potential \end{tabular}  & \begin{tabular}[c]{@{}c@{}}6 layer\\ potential\end{tabular}  & \begin{tabular}[c]{@{}c@{}} Previous \\works \end{tabular} \\
\hline
\begin{tabular}[c]{@{}c@{}} Stacking fault\\ Bernal \end{tabular} & 0                 & 0.10   &  \begin{tabular}[c]{@{}c@{}} 0.14 (RPA) \cite{Wang2015}           \\ 0.14 (Exp.)\cite{Telling2003} \end{tabular} \\
\hline \\
Small barrier $V_\textrm{SP}$    & 1.58              & 1.59   & 1.53 (RPA) \cite{Zhou2015}           \\
\hline \\
Large barrier $V_\textrm{AA}$    & 9.7	               & 9.5  & \begin{tabular}[c]{@{}c@{}}8.8 (RPA) \cite{Zhou2015} \\ 12.4 (QMC) \cite{Mostaani2015} \end{tabular}             \\
\hline \hline \\
\end{tabular}} \\
\end{tabCaption}
\label{tab:energies}
\end{table*}

\clearpage


\begin{thebibliography}{400}
\bibitem{Pierucci2015} D. Pierucci, H. Sediri, M. Hajlaoui, J.-C. Girard, T. Brumme, M. Calandra, E. V.-Fort, G. Patriarche, M. Silly, G. Ferro, V. Souliere, M. Marangolo, F. Sirotti, F. Mauri, and A. Ouerghi, Evidence for Flat Bands Near Fermi Level in Epitaxial Rhombohedral Multilayer Graphene, ACS Nano2015 \textbf{9}, 5432-5439 (2015).

\bibitem{Pamuk2017} Bet\"{u}l Pamuk, J. Baima, F. Mauri, and M. Calandra, Magnetic gap opening in rhombohedral-stacked multilayer graphene from first principles, Phys. Rev. B \textbf{95}, 075422 (2017).

\bibitem{Kopnin2013} N. B. Kopnin, M. Ij\"{a}s, A. Harju, and T. T. Heikkil\"{a}, High-temperature surface superconductivity in rhombohedral graphite, Phys. Rev. B \textbf{87}, 140503 (2013).

\bibitem{Munioz2013} W. Munioz, L. Covaci, and F. Peeters, Tight-binding description of intrinsic superconducting correlations in multilayer graphene, Phys. Rev. B \textbf{87}, 134509 (2013).

\bibitem{Baima2018} J. Baima, F. Mauri, and M. Calandra, Field-effect-driven half-metallic multilayer graphen Phys. Rev. B \textbf{98}, 075418 (2018).

\bibitem{Koshino2010} M. Koshino, Interlayer screening effect in graphene multilayers with ABA and ABC stacking, Phys. Rev. B \textbf{81}, 125304 (2010)	

\bibitem{Xiao2011} R. Xiao, F. Tasnadi, K. Koepernik, J. Venderbos, M. Richter, M. Taut, Phys. Rev. B: Condens. Matter Mater. Phys. \textbf{84}, 165404 (2011).

\bibitem{Laves1956} F. Laves and Y. Baskin, On the Formation of the Rhombohedral Graphite Modification, Zeitschrift für Kristallographie \textbf{107}, 337—356 (1956).

\bibitem{Roviglione2013} Alicia N. Roviglione and Jorge D. Hermida, Rhombohedral Graphite Phase in Nodules from Ductile Cast Iron, Procedia Materials Science \textbf{8},  924 – 933 (2015).

\bibitem{Xu2015} R. Xu, L.-J. Yin, J.-B. Qiao, K.-K. Bai, J.-C. Nie, L. He, Phys. Rev. B \textbf{91}, 035410 (2015).

\bibitem{Kim2015} D.-S. Kim, , H. Kwon, A. Y. Nikitin, S. Ahn, L. Martin-Moreno, F. J. Garcia-Vidal, S. Ryu, H. Min, and Z. H. Kim, ACS Nano 2015 \textbf{9}, 6765−6773 (2015).

\bibitem{Herrero2018} Y. Cao, V. Fatemi, S. Fang, K. Watanabe, T. Taniguchi, E. Kaxiras, and P. Jarillo-Herrero, Unconventional superconductivity in magic-angle graphene superlattices, Nature volume \textbf{556}, 43-50 (2018).

\bibitem{Yankowitz2019} Matthew Yankowitz1, Shaowen Chen Hryhoriy Polshyn, Yuxuan Zhang, K. Watanabe, T. Taniguchi, David Graf, Andrea F. Young, Cory R. Dean, Tuning superconductivity in twisted bilayer graphene, \textbf{363}, 1059-1064 (2019).

\bibitem{Lu2019} Xiaobo Lu, Petr Stepanov, Wei Yang, Ming Xie, Mohammed Ali Aamir, Ipsita Das, Carles Urgell, Kenji Watanabe, Takashi Taniguchi, Guangyu Zhang, Adrian Bachtold, Allan H. MacDonald and  Dmitri K. Efetov, Superconductors, orbital magnets and correlated states in magic-angle bilayer graphene, Nature \textbf{574}, 653–657 (2019).

\bibitem{Lipson1942} H. Lipson and A. R. Stokes, Proc. Roy. Sot. (London) \textbf{A181}, 101 (1942).

\bibitem{Boehm1955} H. P. Boehm and U. Hofmann, Die rhomboedrische Modifikation des Graphites. Ζ. anorg. allg. Chem. \textbf{278}, 58—77 (1955).

\bibitem{Lin2012} Q. Y. Lin, T. Q. Li, Z. J. Liu, Y. Song, L. L. He, Z. J. Hu, Q. G. Guo, H. Q. Ye, High-resolution TEM observations of isolated rhombohedral
crystallites in graphite blocks, Carbon \textbf{50}  2347 – 2374 (2012).

\bibitem{Balseiro2016} Y. Henni, H. P. O. Collado, K. Nogajewski, M. R. Molas, G. Usaj, C. A. Balseiro, M. Orlita, M. Potemski, and C. Faugeras, Rhombohedral Multilayer Graphene: A Magneto-Raman Scattering Study, Nano Lett. \textbf{16}, 3710−3716 (2016).

\bibitem{Henck2018} H. Henck, J. Avila, Z. B. Aziza, D. Pierucci, J. Baima, B. Pamuk, J. Chaste, D. Utt, M. Bartos, K. Nogajewski, B. A. Piot, M. Orlita, M. Potemski, M. Calandra, M. C. Asensio, F. Mauri, C. Faugeras, and A. Ouerghi, Flat electronic bands in long sequences of rhombohedral-stacked graphene, Phys. Rev. B \textbf{97}, 245421 (2018).
 
\bibitem{Mishchenko2019a} Yang, Y. et al, Stacking Order in Graphite Films Controlled by van der Waals Technology, Nano Lett. \textbf{19}, 8526-8532 (2019).

\bibitem{Mishchenko2019b}Y. Shi et. al., Electronic phase separation in topological surface states of rhombohedral graphite, arXiv:1911.04565 (2019).

\bibitem{Dienwiebel2004} M. Dienwiebel, G. S. Verhoeven, N. Pradeep, and J. W. M. Frenken, J. A. Heimberg, H. W. Zandbergen, Superlubricity of Graphite, Phys. Rev. Lett. \textbf{92}, 126101 (2004).

\bibitem{Liu2012b} Z. Liu, S.-M. Zhang, J.-R. Yang, J. Z. Liu, Y.-L. Yang, Q.-S. Zheng, Interlayer shear strength of single crystalline graphite, Acta Mechanica Sinica \textbf{28} 978-982 (2012).

\bibitem{Charlier1994a} J.-C. Charlier, X. Gonze, and J.-P. Michenaud, First-principles study of the stacking effect on the electronic properties of graphite, Carbon \textbf{32}, pp. 289-299 (1994).

\bibitem{Lebedeva2011} I. V. Lebedeva, A. V. Lebedev, A. M. Popov, and A. A. Knizhnik, Comparison of performance of van der Waals-corrected exchange-correlation functionals for interlayer interaction in graphene and hexagonal boron nitride, Computacional Material Science \textbf{128}, 45-58 (2017).

\bibitem{Mounet2005} N. Mounet and N. Marzari, First-principles determination of the structural, vibrational and thermodynamic properties of diamond, graphite, and derivatives, Phys. Rev. B \textbf{71}, 205214 (2005).

\bibitem{Zhou2015} S. Zhou, J. Han, S. Dai, J. Sun, D.J. Srolovitz, Van der Waals bilayer energetics:
Generalized stacking-fault energy of graphene, boron nitride, and graphene boron nitride bilayers, Phys. Rev. B \textbf{92}, 155438 (2015).

\bibitem{Savini2011} G. Savini, Y.J. Dappe , S. \"Oberg, J.-C. Charlier, M.I. Katsnelson, and A. Fasolino, Bending modes, elastic constants and mechanical stability of graphitic systems, Carbon \textbf{49}, 62-69 (2011).

\bibitem{Verhoeven2004} G. S. Verhoeven, M. Dienwiebel, and J. W. M. Frenken, Model calculations of superlubricity of graphite, PHYSICAL REVIEW B \textbf{70}, 165418 (2004).

\bibitem{Wijn2010} A. S. de Wijn, C. Fusco, and A. Fasolino, Stability of superlubric sliding on graphite, Phys. Rev. E \textbf{81}, 046105 (2010).

\bibitem{QE} P. Giannozzi, S. Baroni, N. Bonini, M. Calandra, R. Car, C. Cavazzoni, D. Ceresoli, G. L. Chiarotti, M. Cococcioni, I. Dabo, A. Dal Corso, S. Fabris, G. Fratesi, S. de Gironcoli, R. Gebauer, U. Gerstmann, C. Gougoussis, A. Kokalj, M. Lazzeri, L. Martin-Samos, N. Marzari, F. Mauri, R. Mazzarello, S. Paolini, A. Pasquarello, L. Paulatto, C. Sbraccia, S. Scandolo, G. Sclauzero, A. P. Seitsonen, A. Smogunov, P. Umari, R. M. Wentzcovitch, J.Phys.:Condens.Matter \textbf{21}, 395502 (2009);
P Giannozzi, O Andreussi, T Brumme, O Bunau, M Buongiorno Nardelli, M Calandra, R Car, C Cavazzoni, D Ceresoli, M Cococcioni, N Colonna, I Carnimeo, A Dal Corso, S de Gironcoli, P Delugas, R A DiStasio Jr, A Ferretti, A Floris, G Fratesi, G Fugallo, R Gebauer, U Gerstmann, F Giustino, T Gorni, J Jia, M Kawamura, H-Y Ko, A Kokalj, E Küçükbenli, M Lazzeri, M Marsili, N Marzari, F Mauri, N L Nguyen, H-V Nguyen, A Otero-de-la-Roza, L Paulatto, S Poncé, D Rocca, R Sabatini, B Santra, M Schlipf, A P Seitsonen, A Smogunov, I Timrov, T Thonhauser, P Umari, N Vast, X Wu and S Baroni, J.Phys.:Condens.Matter \textbf{29}, 465901 (2017).

\bibitem{Sabatini2013} R. Sabatini, T. Gorni, and S. de Gironcoli, Nonlocal van der Waals density functional made simple and efficient, Phys. Rev B \textbf{87}, 041108(R) (2013).

\bibitem{Grande2019} R. R. Del Grande, M. G Menezes and, R. B. Capaz, Layer breathing and shear modes in multilayer graphene: a DFT-vdW study, J. Phys.: Condens. Matter \textbf{31} 295301 (2019).

\bibitem{Wang2015} W. Wang, S. Dai, X. Li, J. Yang, D. J. Srolovitz, and Q. Zheng, Measurement of the cleavage energy of graphite, Nat. Comm. \textbf{6}, 7853 (2015).

\bibitem{Telling2003} R. H. Telling and M. I. Heggie, Stacking fault and dislocation glide on the basal
plane of graphite, Philosophical Magazine Letters \textbf{83}, 411–421 (2003).

\bibitem{Baker1961} C.  Baker, Y. T. Chou, and A. Kelly,  Phil. Mag. \textbf{6}, 1305 (1961).

\bibitem{Amelinckx1965} S. Amelinckx, P. Delavignette, and M. Heerschap, Chem. Phys. Carbon \textbf{1}, 1 (1965).

\bibitem{Mostaani2015} E. Mostaani, N. D. Drummond, V. I. Fal’ko, Quantum Monte Carlo calculation of the binding energy of bilayer graphene, Phys. Rev. Lett. \textbf{115}, 115501 (2015).


\end{thebibliography}
\end{document}